\documentclass[cameraready]{Interspeech}

\title{Affect Decoding in Phonated and Silent Speech Production from Surface EMG}

\author[affiliation={1,5}, orcid=0009-0002-6763-7981, equalcontribution]{Simon}{Pistrosch}
\author[affiliation={2}, orcid=0000-0003-0308-795X, equalcontribution]{Kleanthis}{Avramidis}
\author[affiliation={3}, orcid=0000-0003-0707-5016]{Zhao}{Ren}
\author[affiliation={2}, orcid=0000-0002-2053-9068]{Tiantian}{Feng}
\author[affiliation={2}, orcid=0009-0009-3033-2042]{Jihwan}{Lee}
\author[affiliation={1,4}, orcid=0009-0008-9188-058X]{Monica}{Gonzalez-Machorro}
\author[affiliation={1,5}, orcid=0000-0001-7151-8526]{Anton}{Batliner}
\author[affiliation={3}, orcid=0000-0002-9809-7028]{Tanja}{Schultz}
\author[affiliation={2}, orcid=0000-0002-1052-6204]{Shrikanth}{Narayanan}
\author[affiliation={1,4,5}, orcid=0000-0002-6478-8699]{Björn W.}{Schuller}

\address{
    $^1$ CHI -- Chair of Health Informatics, TUM University Hospital, Munich, Germany \\
    $^2$ SAIL -- Signal Analysis and Interpretation Lab, University of Southern California, USA \\
    $^3$ CSL -- Cognitive Systems Lab, University of Bremen, Germany \\
    $^4$ GLAM -- Group on Language, Audio, \& Music, Imperial College London, UK \\
    $^5$ MCML -- Munich Center for Machine Learning, Germany
}

\email{simon.pistrosch@tum.de, avramidi@usc.edu}


\usepackage{comment}

\begin{document}

\maketitle

\begin{abstract}
The expression of affect is integral to spoken communication, yet, its link to underlying articulatory execution remains unclear. Measures of articulatory muscle activity such as EMG could reveal how speech production is modulated by emotion alongside acoustic speech analyses.  We investigate affect decoding from facial and neck surface electromyography (sEMG)  during phonated and silent speech production. For this purpose, we introduce a dataset comprising 2,780 utterances from 12 participants across 3 tasks, on which we evaluate both intra- and inter-subject decoding using a range of features and model embeddings. Our results reveal that EMG representations reliably discriminate frustration with up to 0.845 AUC, and generalize well across articulation modes. Our ablation study further demonstrates that affective signatures are embedded in facial motor activity and persist in the absence of phonation, highlighting the potential of EMG sensing for affect-aware silent speech interfaces.
\end{abstract}

\section{Introduction}

Affect modulation is a central component of spoken communication. Beyond conveying lexical intent, the speech signal encodes paralinguistic information such as attitude, politeness, frustration, and other emotional states \cite{schuller2014computational}. These affective cues are expressed through coordinated muscle activations leading to changes in prosody, articulation, and speech timing across the facial, laryngeal, and respiratory systems. Acoustic analysis has traditionally been used to study affect in speech \cite{schuller2014computational}, as the acoustic signal reflects the downstream consequences of underlying motor control. Understanding how affect is embedded within the motor execution of speech itself remains an important and relatively underexplored question.

Robust affect decoding is particularly relevant in settings where acoustic information is limited, distorted, or unavailable. In assistive communication technologies, silent speech interfaces, and speech prostheses, the ability to recover not only lexical content but also affective intent is crucial for natural and expressive communication \cite{denby2010silent}. Similarly, in atypical speech conditions—such as motor speech disorders \cite{pell2006impact}, post-laryngectomy speech \cite{haderlein2007automatic}, or low-audibility environments—acoustic cues may not reliably convey emotional nuance \cite{schuller2006emotion}. Thus, sensing peripheral speech motor activity may provide an alternative pathway to estimate both speech content and affective state \cite{meltzner2017silent}.

\begin{figure}[t]
  \centering
  \includegraphics[width=\linewidth]{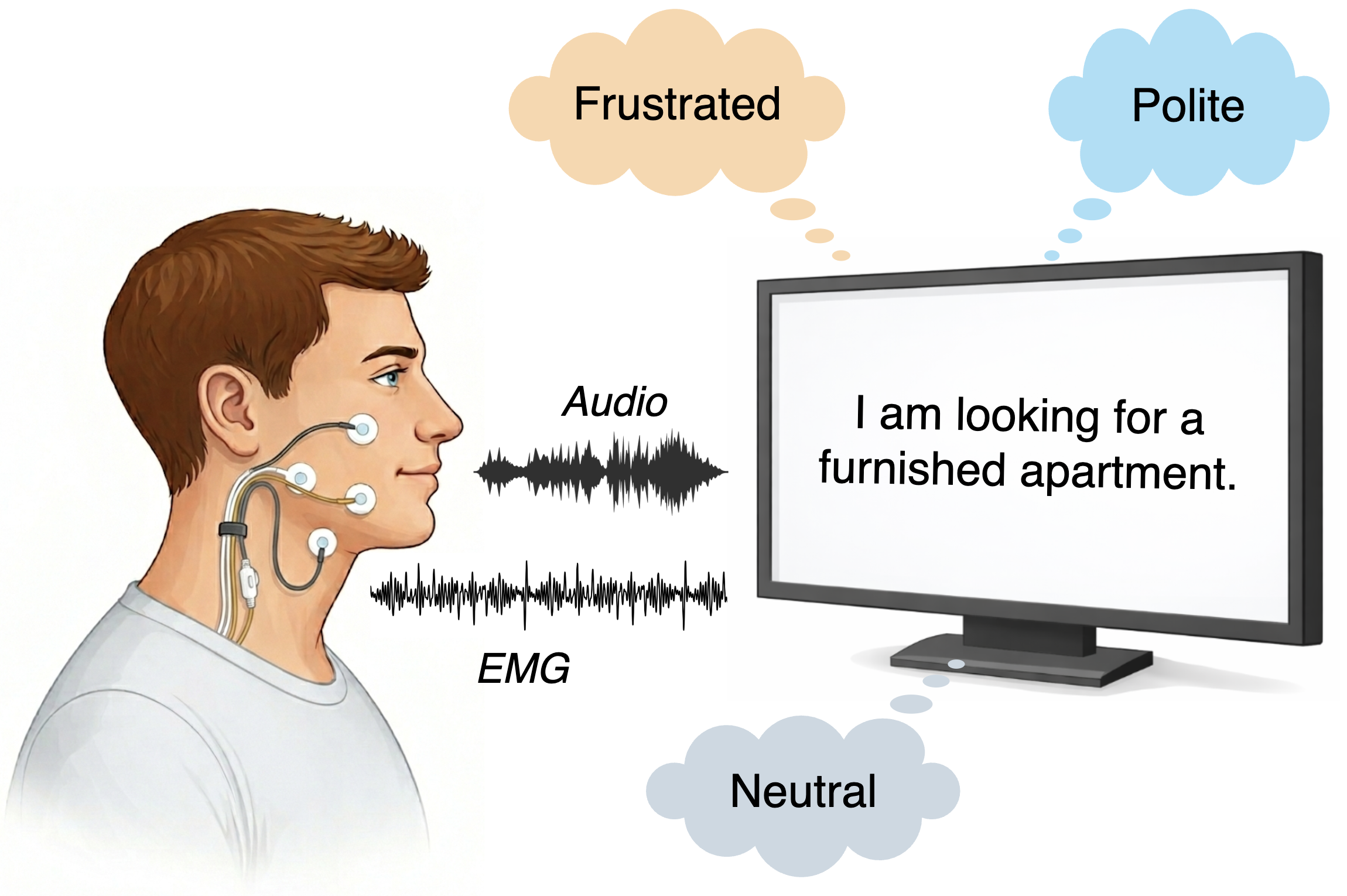}
  \caption{\textbf{Conceptual overview of the study.} We present a dataset and computational analysis on EMG-based affect decoding during phonated and silent speech production. During articulation, surface EMG from neck and facial muscles was recorded alongside audio speech. Note: The schematic human illustration was generated with AI assistance for visualization purposes and is not meant to reflect the exact sensor hardware design, number of channels, or placement used in the study.}
\end{figure}

Surface electromyography (sEMG) offers a non-invasive method for measuring muscle activity underlying speech production. Prior work has demonstrated that articulatory and neck EMG signals can be used to decode phonetic content, recognize silently mouthed words, and even reconstruct phonated speech \cite{vojtech2021surface,scheck2024cross, lee25d_interspeech}. These advances have positioned sEMG as a promising approach for silent speech recognition and speech prosthesis applications \cite{gonzalez2020silent}. As sEMG captures peripheral motor execution, it provides access to the neuromuscular processes underlying speech production, potentially revealing aspects that are not fully observable in the acoustic domain.

However, most EMG-to-speech research has focused on recovering linguistic content rather than characterizing paralinguistic or affective modulation \cite{janke2017emg,gaddy2020digital}. In parallel, facial EMG has been widely used to study emotion-related muscle activity, typically in passive paradigms involving reactions to emotional stimuli \cite{rutkowska2024optimal,jerritta2014emotion}. Relatively little work has examined how affective state modulates the structured motor programs of speech production, particularly across articulation modes such as phonated and silent speech, or under varying recording conditions \cite{ren2025introduction,diener2020towards}. As a result, it remains unclear whether affective signatures are reliably encoded in peripheral muscles and how robust such signatures are across speakers and contexts.

\textbf{Contributions\,} This study systematically investigates affective modulation of speech production as reflected in peripheral muscle activity recorded from neck and facial sEMG. We are particularly interested in how affect is embedded in motor execution across articulation modes, speakers, and recording contexts. We address the following research questions:

\begin{itemize}
    \item \textbf{RQ1:} \emph{To what extent can affective state be decoded from surface EMG during speech production?} 
    We evaluate affect prediction from facial and neck EMG signals during phonated speech, and establish whether peripheral motor activity carries discriminative affective signatures.
    
    \item \textbf{RQ2:} \emph{How does affective state decoding differ between phonated and silent speech production?} 
    We compare decoding performance and signal characteristics across articulation modes to determine whether affective modulation persists in the absence of overt phonation and acoustic output.
    
    \item \textbf{RQ3:} \emph{How does the experimental context influence affective motor signatures?} 
    We examine differences between controlled and spontaneous speech settings to assess the robustness and ecological validity of affective state decoding.
    
\end{itemize}

\section{Related Work}

\subsection{EMG-based speech recognition}

Surface electromyography (sEMG) has been widely studied for silent speech recognition and EMG-to-speech reconstruction \cite{gonzalez2020silent}. Prior work demonstrates that articulatory and neck EMG signals can be used to decode phonetic content, words, and even reconstruct intelligible speech in both phonated and silent articulation settings \cite{scheck2024cross,gaddy2020digital,diener2015direct}. These studies primarily focus on recovering linguistic content or acoustic waveforms, often addressing transfer between phonated and silent speech or improving recognition robustness. In contrast, our work does not aim to reconstruct lexical content. Instead, we investigate whether affective state modulates speech motor execution and whether such modulation can be decoded from peripheral muscle activity across phonated and silent speech production.

\subsection{Affect decoding from facial EMG}

A separate body of research has examined affect detection from facial EMG, typically measuring activity from muscles such as the corrugator supercilii and zygomaticus major \cite{rutkowska2024optimal,tan2016recognition,sato2013relationships}. These studies often rely on passive emotion elicitation paradigms (e.g., emotional images or videos) and analyze spontaneous facial expressions associated with valence or arousal \cite{rymarczyk2016dynamic}. While this literature establishes that affective states are reflected in facial muscle activity, it generally does not consider speech production as a structured motor act under linguistic constraints~\cite{topolinski2014oral}. In speech, affective modulation must operate within already defined articulatory and prosodic variations~\cite{lee2005articulatory,busso2007interrelation}. Our work therefore examines affect not as a standalone facial expression, but as modulation embedded within controlled and spontaneous speech production.

\subsection{Robustness across speakers and contexts}

Relatively little work has investigated affect decoding from speech-related EMG under varying generalization conditions, such as cross-speaker transfer, cross-session robustness, or differences between controlled and spontaneous speech~\cite{scheck2024cross}. Silent speech recognition studies often focus on subject-dependent performance~\cite{gaddy2020digital}, and facial EMG emotion studies frequently emphasize intra-subject analyses~\cite{kolodzieji2024acquisition}. The extent to which affective modulation of speech motor execution generalizes across speakers and articulation modes remains underexplored. By evaluating affect decoding across multiple speakers and across controlled and spontaneous speech conditions, our study contributes to understanding the stability and transferability of affective motor signatures in peripheral speech musculature.

\section{Data Collection}

To address our research questions, we designed a controlled yet ecologically grounded speech production protocol that systematically varied affective state, articulation mode (phonated vs. silent), and interaction context (scripted vs. spontaneous). The resulting dataset comprises multimodal recordings of facial and neck sEMG, as well as audio recordings.

\subsection{Participant recruitment}

Participants were recruited from the city of Munich through fliers that were distributed across the TUM University Hospital, TUM and LMU campuses, as well as in other public places within the centre of Munich. Participants received 12 EUR per hour for their participation. The recruitment materials were approved by the Ethics Committee of the University of Augsburg. The fliers prompted prospective participants to complete an interest form, through which they indicated their demographic information and answered questions to assess eligibility. To qualify for the study, individuals needed to (1)~be at least 18 years old, (2)~not have any current psychiatric or neurological diagnoses, (3)~have normal or corrected-to-normal hearing, and (4)~be native in English or have obtained a C2-equivalent degree. The last qualification was required because the experimental task, which also included production of spontaneous speech, was designed in English.
The dataset could not be made publicly available to protect the pariticpants' privacy. However, data may be shared with third parties if appropriate safeguards are in place according to Article 46 of the General Data Protection Regulation (GDPR). These safeguards include the execution of the European Commission's Standard Contractual Clauses (SCCs).

\subsection{Experimental protocol}

Eligible participants then scheduled an in-person meeting to complete the recording. The experimental session required approximately two hours per participant, comprising one hour of preparation and one hour of active recording. The participants were instructed before the recording day to shave appropriately and avoid makeup, so that the relevant positions for electrode placement are accessible. At the beginning of the experiment, participants were briefed on the protocol and gave informed consent. Subsequently, they completed a detailed demographic questionnaire, which included age, gender, height, weight, and sociolinguistic (dialect/accent) and sociodemographic data as well as the Ten-Item Personality Inventory (TIPI; Gosling et al.~\cite{gosling2003averybrief}) survey. Initially, 15 participants were recruited for the study. Two participants were excluded from analyses due to severely distorted signal recordings, and one withheld consent for data sharing, resulting in a final sample of 12 participants.

The experiment was divided into three tasks. The first was a \textit{prompted reading task}, in which participants were asked to read simple sentences displayed on the screen. While the sentences were generally unrelated, they were purposely selected within a conversational context of \textit{apartment search} (example sentences are displayed in Table~\ref{tab:task1-samples}). There were 50 sentences in total, spanning three different affective states: neutral, polite, and frustration. Each trial directed the participants to articulate the sentence in one of those three ways. The sentences were presented in the following temporal order:

\begin{enumerate}
    \item 10 neutral sentences in a neutral way
    \item 10 polite-worded sentences in a polite way
    \item 10 neutral sentences (same as the first 10) in a polite way
    \item 10 frustration-worded sentences in a frustrated way
    \item 10 neutral sentences (same as the first 10)  in a frustrated way
\end{enumerate}

\begin{table*}[t]
    \caption{Example sentences from the prompted reading task (apartment search). Neutral sentences were articulated in three expressive tones (neutral, polite, frustrated), whereas polite- and frustration-worded sentences were spoken congruently with their lexical content.}
    \label{tab:task1-samples}
    \centering
    \begin{tabular}{p{8.5cm} p{3.2cm} p{3.2cm}}
        \toprule
        \textbf{Utterance} & \textbf{Lexical category} & \textbf{Expressed tone(s)} \\
        \midrule
        I am looking for a furnished apartment. 
        & Neutral 
        & Neutral, polite, frustrated \\
        I would be delighted if you had a furnished apartment available. 
        & Polite-worded 
        & Polite \\
        Why isn't the landlord responding to my inquiry? 
        & Frustration-worded 
        & Frustrated \\
        \bottomrule
    \end{tabular}
\end{table*}

\begin{table*}[t]
\caption{Dataset statistics by task and speaking mode. Utt. denotes number of utterances; Mins denotes total minutes. Recording duration is shown in seconds (s).}
\label{tab:dataset_stats}
\centering
\begin{tabular}{l ccc ccc ccc}
\toprule
\multirow{2}{*}{\textbf{Task}} 
& \multicolumn{3}{c}{\textbf{Phonated}} 
& \multicolumn{3}{c}{\textbf{Silent}} 
& \multicolumn{3}{c}{\textbf{Combined}} \\
\cmidrule(lr){2-4} \cmidrule(lr){5-7} \cmidrule(lr){8-10}
& \textbf{Utt.} & \textbf{Mins} & \textbf{Per recording (s)} 
& \textbf{Utt.} & \textbf{Mins} & \textbf{Per recording (s)} 
& \textbf{Utt.} & \textbf{Mins} & \textbf{Per recording (s)} \\
\midrule

Task 1
& 593 & 36.6 & 3.70 $\pm$ 1.09 
& 592 & 38.3 & 3.88 $\pm$ 1.21 
& 1185 & 74.9 & 3.79 $\pm$ 1.15 \\

Task 2
& 396 & 47.5 & 7.20 $\pm$ 6.14 
& -- & -- & -- 
& 396 & 47.5 & 7.20 $\pm$ 6.14 \\

Task 3
& 599 & 36.3 & 3.63 $\pm$ 1.07 
& 600 & 35.5 & 3.55 $\pm$ 1.02 
& 1199 & 71.8 & 3.59 $\pm$ 1.05 \\

\midrule

Total 
& 1588 & 120.4 & 4.55 $\pm$ 3.55 
& 1192 & 73.9 & 3.72 $\pm$ 1.13 
& 2780 & 194.2 & 4.19 $\pm$ 2.81 \\

\bottomrule
\end{tabular}
\end{table*}

\begin{figure}
    \centering
    \includegraphics[width=\linewidth]{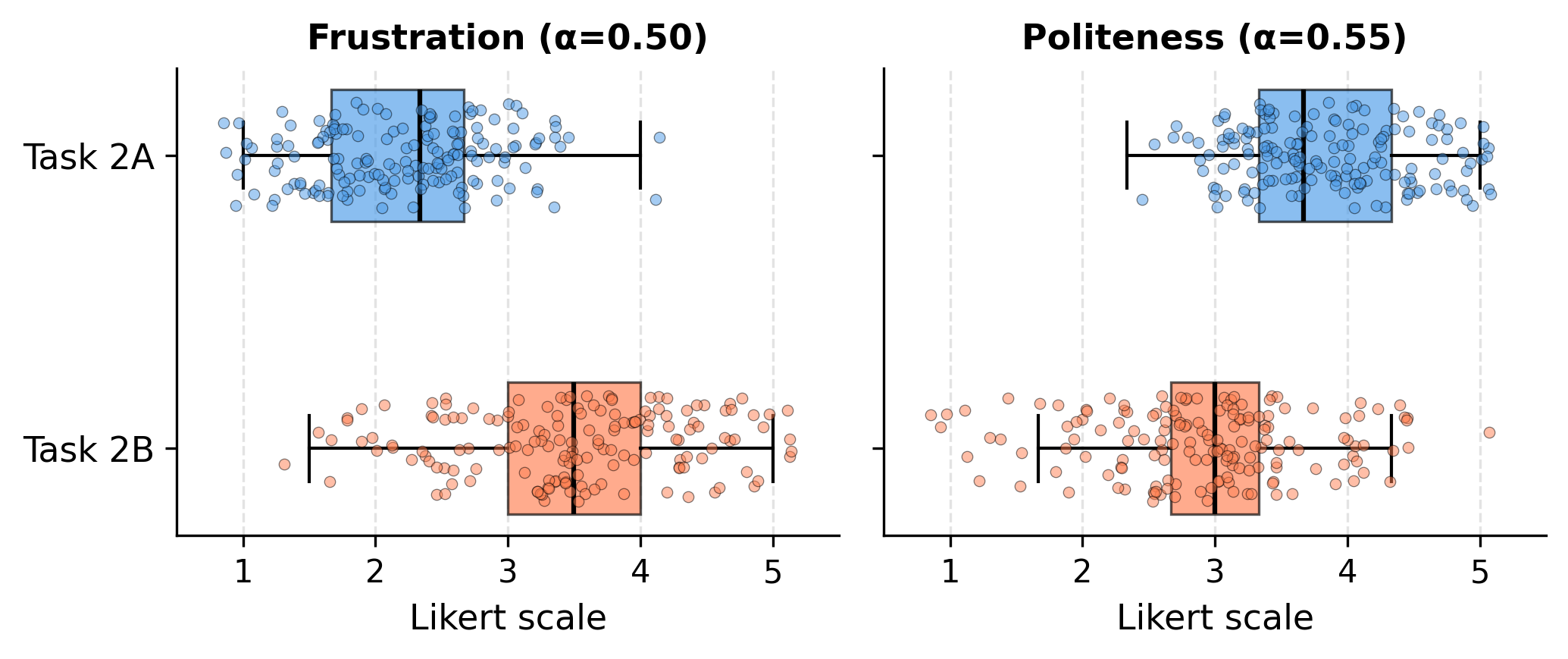}
    \caption{Annotation results for Task 2A (designed to induce politeness) and Task 2B (designed to induce frustration). Individual trial annotations are overlaid to the boxplots, pooled across the 3 annotators. Inter-annotator agreement is included in terms of Krippendorff's alpha. Light jittering is applied to the integer annotation values for visualization purposes.}
    \label{fig:annotations}
    \vspace{-0.1cm}
\end{figure}

Moreover, each sentence was prompted twice; the first time the participants were directed to phonate the sentence aloud, whereas the second time, immediately following the first, to do it silently without vocalizing. Therefore, Task 1 resulted in 100 trials per participant. Task 3, conducted after about 30 minutes, was an exact repetition of the first task.

For the second task, participants were asked to \textit{spontaneously converse} with a Wizard-of-Oz agent. The conversational context was a \textit{car insurance discussion}. Participants were given a sheet with all information necessary for the task, including a fake name, and were asked to only speak audibly. The task was further divided into two scenarios; the first was designed to naturally elicit polite responses, whereas the second to elicit frustration. While the responses of the participant were spontaneous, the topics discussed were controlled and the agent's responses were selected by a member of the experimental team.

In the first session, participants were required to request a bonus to reduce their insurance costs, register a second vehicle, and apply for a green card for driving abroad. In the second session, participants were instructed to contact the insurance company to request payment of an outstanding quotation. The agents' speech was synthesized using ElevenLabs (elevenlabs.io). The polite agent employed a warm, friendly prosody and formal language. The frustration agent used a harsher vocal tone and abrupt, impolite language. Additionally, this agent simulated poor comprehension, frequently requesting repetitions. Participants were blinded to the Wizard-of-Oz methodology, operating under the premise that they were interacting with an autonomous ChatGPT agent. They remained unaware that the two agents were explicitly designed to elicit politeness and frustration. Upon completing both tasks, participants were fully debriefed regarding the experimental setup.

\subsection{Sensor Apparatus}

All recordings took place at the TUM University Hospital, and were conducted in an acoustically treated room. The recording conditions were optimized using wall-mounted acoustic panels, bass traps, and curtains to suppress reverberation and minimize ambient noise. The EMG signal was recorded using an actiCHamp Plus (Brain Products GmbH, Gilching, Germany) amplifier with 8 bipolar surface EMG electrodes (Ag/AgCl). The device was connected via USB to the recording computer. Before attaching an electrode to a specific site on the participant’s skin, the area was cleaned with an alcoholic solution and skin preparation gel. Electrode placements are summarized in Table~\ref{tab:electrode-positions-extended}, with the ground electrode placed at the the end of the nasal bone. Neck and facial muscles of interest were determined based on their utility in speech production and affective expression~\cite{gaddy2020digital,parent2002issues,chen2022emotion,lopez2010syllable,wand2014emg,zhang2023emgbased}. Ten20 conductive gel was applied to each electrode, and impedances were kept below \SI{100}{\kilo\ohm}. EMG was sampled at an initial rate of \SI{10}{\kilo\hertz}.


\begin{table*}[t]
\caption{Electrode placement and functional classification of recorded muscles. Speech- and emotion-related assignments are derived from reported functional associations~\cite{gaddy2020digital,parent2002issues,chen2022emotion,lopez2010syllable,wand2014emg,zhang2023emgbased}. A schematic of the described placement is shown in ablation Figure~\ref{fig:ablation1}.}
\label{tab:electrode-positions-extended}
\centering
\small
\begin{tabular}{p{1.0cm} p{3.4cm} p{7.3cm} c c}
\toprule
\textbf{ID} & \textbf{Muscle site} & \textbf{Positioning} & \textbf{Speech} & \textbf{Emotion} \\
\midrule

E1 & 
Infrahyoid &
2.5\,cm lateral to thyroid prominence. Avoid direct proximity to reduce signal contamination. 
& \checkmark & -- \\
E2 & 
Suprahyoid &
Midway between mandible and relaxed hyoid, ~0.5\,cm lateral to midline. 
& \checkmark & -- \\
E3 & 
Mylohyoid &
Midway between chin center and lateral endpoint, ~2\,cm inferior to chin line (submental). 
& \checkmark & -- \\
E4 & 
Mentalis &
0.5\,cm lateral to midline and 0.5\,cm superior to pogonion. 
& \checkmark & \checkmark \\
E5 & 
Orbicularis Oris Superioris &
Facial midline above upper lip. 
& \checkmark & -- \\
E6 & 
Depressor Supercilii &
Between nasal bone and eyebrow. 
& -- & \checkmark \\
E7 & 
Zygomaticus Major (left) &
Caudal end of zygomatic bone along line to lip corner. 
& \checkmark & \checkmark \\
E8 & 
Zygomaticus Major (right) &
Mirrored placement of E7. 
& \checkmark & \checkmark \\
\bottomrule
\end{tabular}
\vspace{-0.1cm}
\end{table*}

Additionally, speech audio was recorded using a Rode NT1-A microphone at a sampling rate of 48 kHz and Focusrite Scarlett 2i2 audio interface. The amplifier and the audio interface were connected to a microprocessor, which transmitted start and end markers for each utterance to all modalities to enable subsequent alignment. For data recording, we adapted the EMG-GUI software originally developed by Diener \cite{diener2021thesis} and extended by Scheck et al. \cite{scheck2025diffmv}. Each trial recording was initiated by the participant pressing a button on the computer screen. The recording lasted for as long as the button was pressed and finished when the participant released the button. All participants were asked to avoid rushing into their response immediately after pressing the button, and to release the button only when they are done speaking. If the participant was not satisfied with a certain trial recording, they were given the option to repeat it as many times as needed.

\subsection{ST-Case dataset}
The ST-CASE (SAIL-TUM Corpus on Affective Speech \& EMG) dataset comprises a sample of $N=12$ participants ($9$ female), five of whom were native English speakers. The mean age of the participants was $26.2$ years ($\text{SD} = 5.2, \text{MIN} = 20,\text{MAX} = 36$). The dataset consists of a total of 2\,780 utterances, divided into 1\,588 phonated and 1\,192 silent recordings. As shown in Table~\ref{tab:label_distribution}, these are further categorized  into 1\,143 frustrated, 479 neutral, and 1\,158 polite utterances (see Table~\ref{tab:dataset_stats} and Table~\ref{tab:label_distribution}). 

The prompted Tasks 1 and 3 followed a distribution where frustrated and polite labels were approximately twice as frequent as neutral labels across both recording modes. The spontaneous Task 2 was limited to the phonated mode and focused on frustration and politeness, contributing $64$ utterances for politeness and $113$ for frustration. For this task, three annotators rated the participants' spontaneous utterances on two 5-point Likert scales (ranging from ``disagree" to ``agree") to assess whether speakers sounded frustrated or polite. The distribution of the annotator scores is shown in Figure~\ref{fig:annotations}. We then averaged the annotators' scores for each scale separately.

Due to the spontaneous nature of Task 2, the average number of utterances per participant was $16.50 \pm 1.88$ for the ``Request of a bonus'' scenario and $16.50 \pm 0.90$ for the ``Unpaid invoice'' scenario. On average, the prompted reading tasks yielded shorter utterance durations compared to spontaneous speech (Task $1$: $3.79$~s; Task $3$: $3.59$~s; Task $2$: $7.20$~s).

\begin{table}[t]
\centering
\caption{Label distribution in terms of number of utterances across tasks and speaking modes. For Task 2, we report on the affect label anticipated by the experimental design.}
\label{tab:label_distribution}
\setlength{\tabcolsep}{5pt}
\begin{tabular}{lcccccc}
\toprule
 & \multicolumn{3}{c}{\textbf{Phonated}} & \multicolumn{3}{c}{\textbf{Silent}} \\
\cmidrule(lr){2-4} \cmidrule(lr){5-7}
\textbf{Task} & \textbf{Frust.} & \textbf{Neut.} & \textbf{Polit.} & \textbf{Frust.} & \textbf{Neut.} & \textbf{Polit.} \\ 
\midrule
Task 1 & 233 & 120 & 240 & 232 & 120 & 240 \\
Task 2 & 198 & -- & 198 & -- & -- & -- \\
Task 3 & 240 & 119 & 240 & 240 & 120 & 240 \\
\midrule
Total & 671 & 239 & 678 & 472 & 240 & 480 \\ 
\bottomrule
\end{tabular}
\end{table}

\subsection{Pre-processing methods}

Raw 8-channel EMG recordings were filtered using a 4th-order high-pass Butterworth filter at \SI{100}{\hertz}, iterative notch filtering at \SI{50}{\hertz} (and up to 8 harmonics), and anti-aliasing low-pass filtering prior to decimation. Signals were downsampled to \SI{1}{\kilo\hertz} with an IIR filter and extreme outliers (±10 standard deviations) were clipped. Each task also includes a baseline recording, which was processed the same way. The baseline was used to compute the median and inter-quartile ratio of the signal at rest, which was subsequently used for robust scaling.

Phonated speech trials are loaded as audio waveforms and denoised using the \texttt{noisereduce} Python package. The signals were then normalized by peak RMS (target RMS was 0.5), and downsampled to \SI{16}{\kilo\hertz}. For the second task, the utterance text is obtained via an automatic Whisper-small transcription and stored in the trial metadata, although this study did not use the transcriptions. We compute conservative onset/offset indices, set as 0.56\,s onset and 0.54\,s margin from the end, applied to both EMG and aligned audio signals.

\section{Methodology}

\subsection{EMG feature extraction}

We extracted a set of handcrafted features and TD-$n$ features~\cite{janke2017emg,diener2020towards} from the EMG signals. Let $x_c \in \mathbb{R}^{T}$ denote the baseline-corrected and $z$-scored EMG signal of channel $c$, and let $\tilde{x}_c[i] = |x_c[i]|$ denote its rectified version. For each channel $c$, we computed the mean rectified value $\overline{\tilde{x}_c}$ along with its standard deviation and coefficient of variation. We also computed the peak amplitude of $\tilde{x}_c$ as well as the root mean square value $\sqrt{\overline{x_c^{2}}}$ of the signal. Additionally, let $\mathrm{PSD}_c(f)$ denote the Welch-estimated power spectral density of $x_c$. We computed the median frequency and spectral entropy of the signal,
\begin{align}
f_{\mathrm{med},c}
&= \frac{1}{2}\int_{0}^{f_{\max}} \mathrm{PSD}_c(f)\,df, \\
H_{\mathrm{spec},c}
&= -\sum_{k} p_{c,k}\log p_{c,k},
\end{align}
where $p_{c,k}$ denotes normalized spectral power. Finally, we derived cross-channel Pearson correlation upon the rectified signals. Each trial was thus encoded into a 92-dimensional vector.

To compute the TD-0 features, the signal is first split into low- and high-frequency components using a triangular filter with a 134~Hz cutoff, implemented via a double moving average. The signal is then segmented into rectangular windows of 27~ms with a 10~ms frame shift. TD features are calculated for each frame with the low- and high-frequency parts and zero-crossing rate (ZCR). For each TD feature, we computed the mean, standard deviation, and the 0th, 25th, 75th, and 100th percentiles across frames. The TD features are defined as:
\begin{multline}
\mathrm{TD}(x_{\mathrm{low}}, x_{\mathrm{high}}) = \Bigg( \frac{1}{n} \sum_{i=1}^{n} (x_{\mathrm{low}}[i])^{2}, \frac{1}{n} \sum_{i=1}^{n} x_{\mathrm{low}}[i], \\
\frac{1}{n} \sum_{i=1}^{n} (x_{\mathrm{high}}[i])^{2}, \frac{1}{n} \sum_{i=1}^{n} |x_{\mathrm{high}}[i]|, \mathrm{ZCR}(x_{\mathrm{high}}) \Bigg).
\end{multline}

To test the applicability of deep learning-based (foundation) models, we also extracted embedding features for our EMG input. We used the BioCodec~\cite{avramidis2025neural} model, which is one of the very few open-source models trained on sEMG signals. Although BioCodec was pre-trained on gesture signals from the wrist, we empirically verified its robustness to our data by inspecting the reconstruction quality of the input. For the purpose of this study, we extracted the output embedding of the BioCodec encoder, i.e., the channel-wise 128D input to the quantization module.

\subsection{Speech feature extraction}

Prosodic features were obtained through the \texttt{eGeMAPSv02} set via openSMILE~\cite{eyben2010opensmile}. We also extracted deep-learning based speech emotion features using Vox-Profile~\cite{feng2025vox}, in the form of 256D embeddings. These were obtained from a dimensional speech emotion model (i.e., trained to estimate arousal, valence, and dominance) that was fine-tuned with the Whisper-Large~\cite{radford2023robust} backbone on the MSP-Podcast dataset~\cite{lotfian2017building}.

\subsection{Machine learning models}

All subsequent experiments were conducted using either handcrafted feature representations or latent model embeddings. Specifically, we trained two classifiers: a Support Vector Machine (SVM) with a radial basis function (RBF) kernel applied to the handcrafted features, and a linear probe classifier with L2 regularization applied to the latent embeddings. Prior to model training, handcrafted features were z-scored using statistics computed from the training set, whereas latent embeddings were scaled using a robust scaler fitted on the training data.

\section{Experimental Setup}

\subsection{RQ1: Affect decoding from sEMG}
\label{sec:rq1}

To investigate whether affective states can be decoded from sEMG during speech production, we conducted both intra-subject and inter-subject evaluations using EMG data from Tasks~1 and~3. In all cases, the objective was to predict the affect label of each individual trial. For the intra-subject analysis, models were trained and evaluated separately for each participant using 5-fold cross-validation over trials. Specifically, trials were partitioned such that all repetitions of a given sentence appeared in the same fold, thereby preventing sentence-level data leakage. For the inter-subject analysis, we employed nested cross-validation. The outer loop followed a leave-one-subject-out (LOSO) scheme to evaluate generalization to unseen participants. Within each training split, we used the same trial-level 5-fold cross-validation over the available data, again ensuring that sentences were not repeated across folds. Model performance was evaluated using the Area-Under-the-Curve (AUC) and balanced accuracy (BAC). The results were reported as mean and standard deviation between subjects for each metric.

\subsection{RQ2: Phonated and silent speech}

We employed the same pre-processing pipeline, modeling and evaluation procedure described in~\ref{sec:rq1}. For this analysis, the dataset (Tasks 1 and 3) was partitioned into two articulation modes: \textit{phonated} and \textit{silent} speech. Each subset was evaluated independently using the same cross-validation strategy as described above. In addition, we investigated cross-setting generalization by training models on one articulation mode (phonated or silent) and evaluating their generalizability on the other.

\subsection{RQ3: From controlled to spontaneous speech}

To investigate the influence of the experimental context, we evaluated model performance on Task~2, containing spontaneous, conversational speech. We employed the same pre-processing pipeline and model configurations as in Section~\ref{sec:rq1}. Because data from Task 2 are relatively scarce (see Table~\ref{tab:dataset_stats}), we trained our models on both the controlled tasks (Tasks~1 and~3) and the spontaneous Task~2, and we share results for the inter-subject configuration, where data from the test speaker were excluded entirely from the training set. Label assignment was based on the conducted data annotations, and specifically on the annotation of \textit{frustration}, as this had a higher agreement with the pre-defined target affect of each sub-task than the annotation of politeness (see Figure~\ref{fig:annotations}). Specifically, we assigned the label of frustration to all trials annotated higher than 3.5, and the label of politeness to all trials annotated lower than 2.5, with the remaining ambiguous trials discarded for clarity.

\begin{table*}[t]
\centering
\caption{Classification performance for affective mode decoding across varying modalities. Results present the mean $\pm$ standard deviation for Balanced Accuracy (BAC) and Area Under the Curve (AUC) across subjects under intra-subject and inter-subject settings.}
\label{tab:rq1}
\begin{tabular}{l l cc cc}
\toprule
\multirow{2}{*}{\textbf{Modality}} &
\multirow{2}{*}{\textbf{Features}} &
\multicolumn{2}{c}{\textbf{Intra-subject}} &
\multicolumn{2}{c}{\textbf{Inter-subject}} \\
\cmidrule(lr){3-4} \cmidrule(lr){5-6}
 & & \textbf{BAC} & \textbf{AUC} & \textbf{BAC} & \textbf{AUC} \\
\midrule

\multirow{3}{*}{EMG} 
& Structural 
& \underline{0.749} $\pm$ 0.075 & \underline{0.820} $\pm$ 0.081 
& 0.546 $\pm$ 0.054 & 0.568 $\pm$ 0.074 \\

& TD-0  
& \textbf{0.762} $\pm$ 0.063 & \textbf{0.845} $\pm$ 0.058 
& 0.541 $\pm$ 0.052 & 0.567 $\pm$ 0.082 \\

& BioCodec  
& 0.721 $\pm$ 0.053 & 0.792 $\pm$ 0.075 
& \underline{0.547} $\pm$ 0.052 & \underline{0.574} $\pm$ 0.080 \\

\midrule

\multirow{2}{*}{Speech}
& eGeMAPS      
& 0.610 $\pm$ 0.125 & 0.644 $\pm$ 0.168 
& 0.527 $\pm$ 0.058 & 0.495 $\pm$ 0.115 \\

& Vox-Profile  
& 0.659 $\pm$ 0.097 & 0.732 $\pm$ 0.104 
& \textbf{0.582} $\pm$ 0.041 & \textbf{0.657} $\pm$ 0.071 \\
\bottomrule
\end{tabular}
\end{table*}

\begin{table}[t]
\centering
\caption{\textbf{Intra-subject} classification performance in terms of mean $\pm$ standard deviation of subject-wise AUC across affectively unique and repeated trials. Repeated trials are those who were encountered in the experiment once for each available label (here, Polite and Frustrated).}
\label{tab:repeated}
\begin{tabular}{l l cc}
\toprule
\textbf{Modality} & \textbf{Features} & \textbf{Unique} & \textbf{Repeated} \\
\midrule

\multirow{3}{*}{EMG}
& Structural  
& \underline{0.856} $\pm$ 0.069 
& 0.720 $\pm$ 0.159 \\
& TD-0   
& 0.824 $\pm$ 0.126 
& \textbf{0.751} $\pm$ 0.145 \\
& BioCodec  
& 0.799 $\pm$ 0.058 
& \underline{0.747} $\pm$ 0.145 \\

\midrule

\multirow{2}{*}{Speech}
& eGeMAPS      
& 0.643 $\pm$ 0.060 
& 0.559 $\pm$ 0.132 \\
& Vox-Profile  
& \textbf{0.889} $\pm$ 0.075 
& 0.469 $\pm$ 0.202 \\

\bottomrule
\end{tabular}
\end{table}

\begin{table}[t]
\centering
\caption{\textbf{Inter-subject} classification performance. Results are reported in terms of mean $\pm$ standard deviation of subject-wise AUC for Task 1 and Task 3 separately. A Wilcoxon signed-rank test was used to determine significant changes between tasks. FDR correction was applied to all p-values.}
\label{tab:inter_auc}
\begin{tabular}{l ccc}
\toprule
\textbf{Features} & \textbf{Task 1} & \textbf{Task 3} & \textbf{p-value} \\
\midrule

Structural  
& 0.527 $\pm$ 0.059 
& \textbf{0.613} $\pm$ 0.100 
& \underline{0.038} \\

TD-0   
& 0.469 $\pm$ 0.121 
& \textbf{0.617} $\pm$ 0.116 
& \underline{0.038} \\

BioCodec  
& 0.520 $\pm$ 0.079 
& \textbf{0.603} $\pm$ 0.112 
& 0.056 \\

\midrule

eGeMAPS      
& \textbf{0.489} $\pm$ 0.126 
& 0.458 $\pm$ 0.113 
& 0.534 \\

Vox-Profile  
& \textbf{0.667} $\pm$ 0.070 
& 0.632 $\pm$ 0.076 
& 0.464 \\

\bottomrule
\end{tabular}
\vspace{-0.1cm}
\end{table}

\begin{figure}[t]
  \centering
  \includegraphics[width=\linewidth]{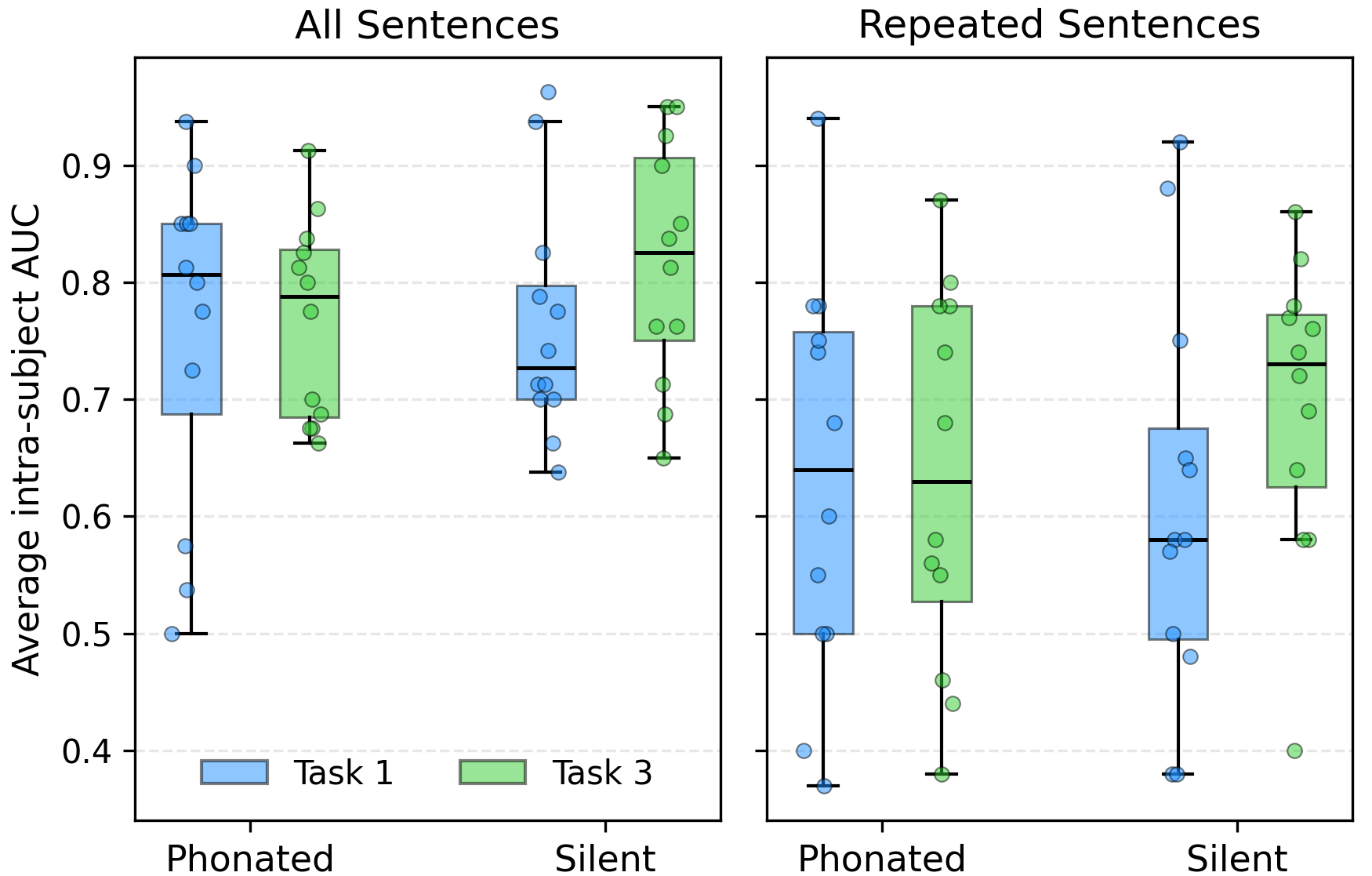}
  \caption{Comparison of intra-subject AUC between Task 1 and Task 3 across speaking conditions. \textbf{Left}: tested on all sentences. \textbf{Right}: tested on the repeated sentences (see also Table~\ref{tab:repeated}). Dots correspond to average individual performance.}
  \label{fig:boxplots}
  \vspace{-0.1cm}
\end{figure}




\begin{table*}[t]
\centering
\caption{Affect decoding performance in terms of AUC (average $\pm$ SD) for two experimental settings and three EMG input sets.
Intra-mode refers to testing within the same articulation mode; cross-mode refers to training on one mode and testing on the other.}
\label{tab:silent_audible}
\begin{tabular}{l l cc cc}
\toprule
\textbf{Setting} & \textbf{Features}
  & \multicolumn{2}{c}{\textbf{Intra-mode}}
  & \multicolumn{2}{c}{\textbf{Cross-mode}} \\
\cmidrule(lr){3-4} \cmidrule(lr){5-6}
 & & Phonated\,$\rightarrow$\,Phonated & Silent\,$\rightarrow$\,Silent
     & Phonated\,$\rightarrow$\,Silent & Silent\,$\rightarrow$\,Phonated \\
\midrule
\multirow{3}{*}{Intra-subject}
  & Structural   & \textbf{0.815} $\pm$ 0.110 & \textbf{0.829} $\pm$ 0.056 & 0.707 $\pm$ 0.158 & 0.663 $\pm$ 0.168 \\
  & TD-0       & 0.806 $\pm$ 0.113 & 0.811 $\pm$ 0.100 & 0.705 $\pm$ 0.118 & 0.626 $\pm$ 0.145 \\
  & BioCodec   & 0.758 $\pm$ 0.066 & 0.792 $\pm$ 0.102 & \textbf{0.763} $\pm$ 0.094 & \textbf{0.745} $\pm$ 0.075 \\
\midrule
\multirow{3}{*}{Inter-subject}
  & Structural     & 0.567 $\pm$ 0.112 & \textbf{0.608} $\pm$ 0.073 & \textbf{0.639} $\pm$ 0.085 & \textbf{0.646} $\pm$ 0.101 \\
  & TD-0       & \textbf{0.589} $\pm$ 0.103 & 0.563 $\pm$ 0.077 & 0.612 $\pm$ 0.071 & 0.644 $\pm$ 0.103 \\
  & BioCodec     & 0.575 $\pm$ 0.075 & 0.592 $\pm$ 0.090 & 0.624 $\pm$ 0.099 & 0.615 $\pm$ 0.095 \\
\bottomrule
\end{tabular}
\end{table*}

\section{Results}

\subsection{RQ1: Reliable affect decoding from EMG}

Table~\ref{tab:rq1} summarizes the binary classification performance for intra-subject and inter-subject settings across EMG and speech modalities, where the neutral trials were excluded to minimize ambiguity. In the intra-subject setting, EMG-based approaches consistently outperformed speech-based features. TD achieves the highest performance (AUC = 0.845), however all three modalities performed similarly, without significant statistical differences. In contrast, speech features yielded more modest results, with eGeMAPS (AUC = 0.644) and Vox-Profile (AUC = 0.732) substantially underperforming the EMG-based models (McNemar $p<0.001$). These findings suggest that prompted affective modulation was strongly expressed through muscle activity, but did not translate to perceived acoustic differences.

Under the more challenging inter-subject classification scenario, overall performance showed regressions across modalities. While EMG-based models yielded comparable results with marginally higher than chance discriminability, Vox-Profile achieved the strongest cross-subject generalization with an AUC of 0.657. These findings verified the large participant heterogeneity when it comes to affective expression, for which a participant cohort of $N=12$ does not suffice to yield global affect markers. We further assume that the higher performance of the speech foundation model was partially attributed to lexical decoding of affective elements in the sentences.

To control for potential lexical confounds, we further evaluated model performance separately for sentences presented only once, with a single affective label (termed \textit{unique}) and for sentences presented twice, each time with a different affective label (termed \textit{repeated}). This separation reduces the possibility that models rely on sentence-specific lexical cues rather than affective expression. The results are reported in Table~\ref{tab:inter_auc}. As expected, all models achieved higher AUC on the unique subset and showed a consistent drop in performance on the repeated sentences, all regressions being statistically significant. Importantly, EMG-based models retain moderate discriminability even in the repeated condition (AUC $>$ 0.7), indicating robustness to lexical overlap. In contrast, speech models degrade substantially: while Vox-Profile achieves the highest performance overall on unique trials (AUC = 0.889), its performance collapses to random-chance levels on repeated sentences (AUC = 0.469). eGeMAPS features exhibit a more modest decline, indicating that prosodic descriptors still captured some affect variation in a controlled lexical context.

\begin{figure}[t]
  \centering
  \includegraphics[width=\linewidth]{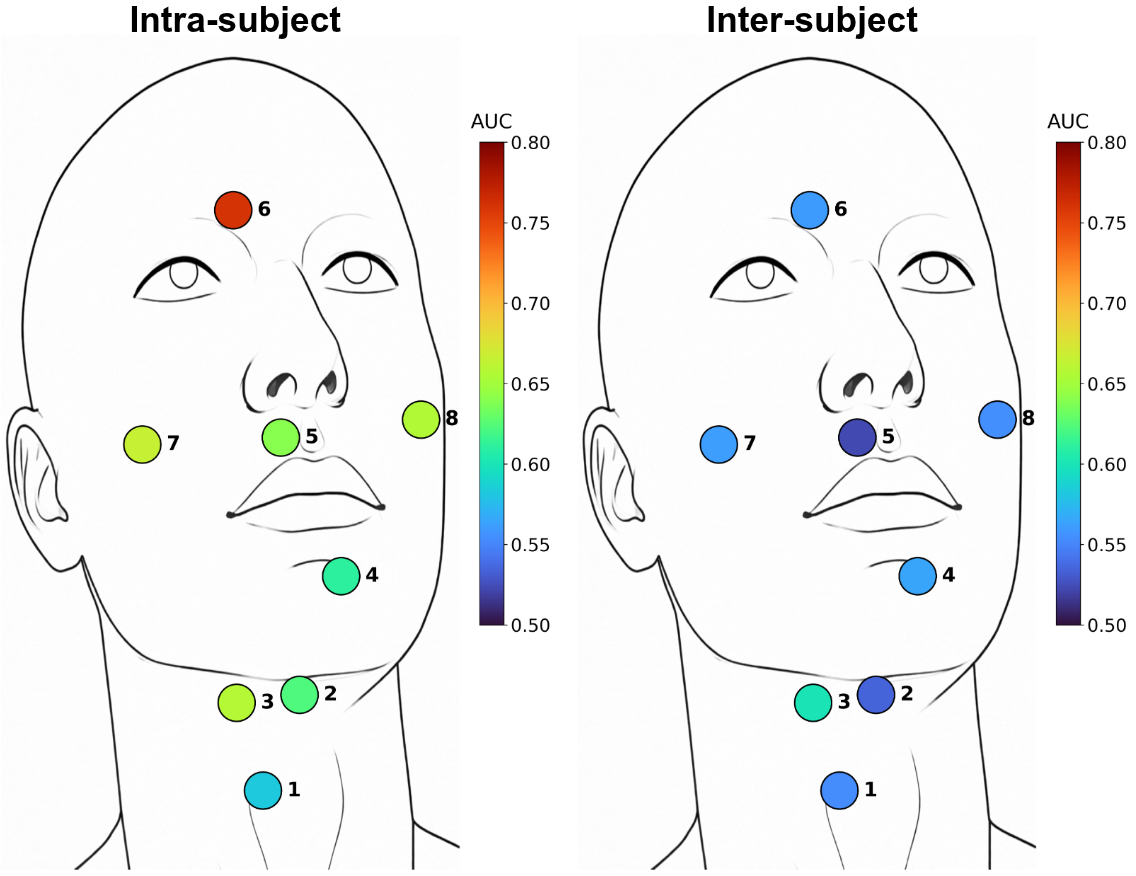}
  \caption{\textbf{Channel-wise decoding performance across evaluation settings (RQ1).} Topographic visualization of electrode-specific AUC for the intra- (left) and inter-subject (right) settings in Tasks 1, 3. Each marker corresponds to an EMG channel, with warmer colors reflecting higher discriminability.}
  \vspace{-0.2cm}
  \label{fig:ablation1}
\end{figure}

Overall, EMG representations demonstrated stronger discriminability than the speech features, indicating that the affective modulation was directed primarily to paralinguistic channels and facial expressions rather than core speech production elements. We tested this assumption through an ablation on the electrode set, and trained separate, single-channel classifiers under identical settings. A topographical heatmap of the results is provided in Figure~\ref{fig:ablation1}, separately for intra-subject (left) and inter-subject (right) scenarios. Several facial channels—most prominently the frontal site (E6)—achieved high discriminability, with additional contributions from perioral and submental locations. Performance appears spatially differentiated, suggesting that affective information is not uniformly distributed across recording sites. In contrast, the inter-subject setting (right) showed regressions in AUC across nearly all channels, indicating limited inter-subject generalization at the individual electrode level. Attenuation was especially pronounced in the frontal and cheek sites, while submental channels retained relatively moderate performance.

\begin{figure}[t]
  \centering
  \includegraphics[width=\linewidth]{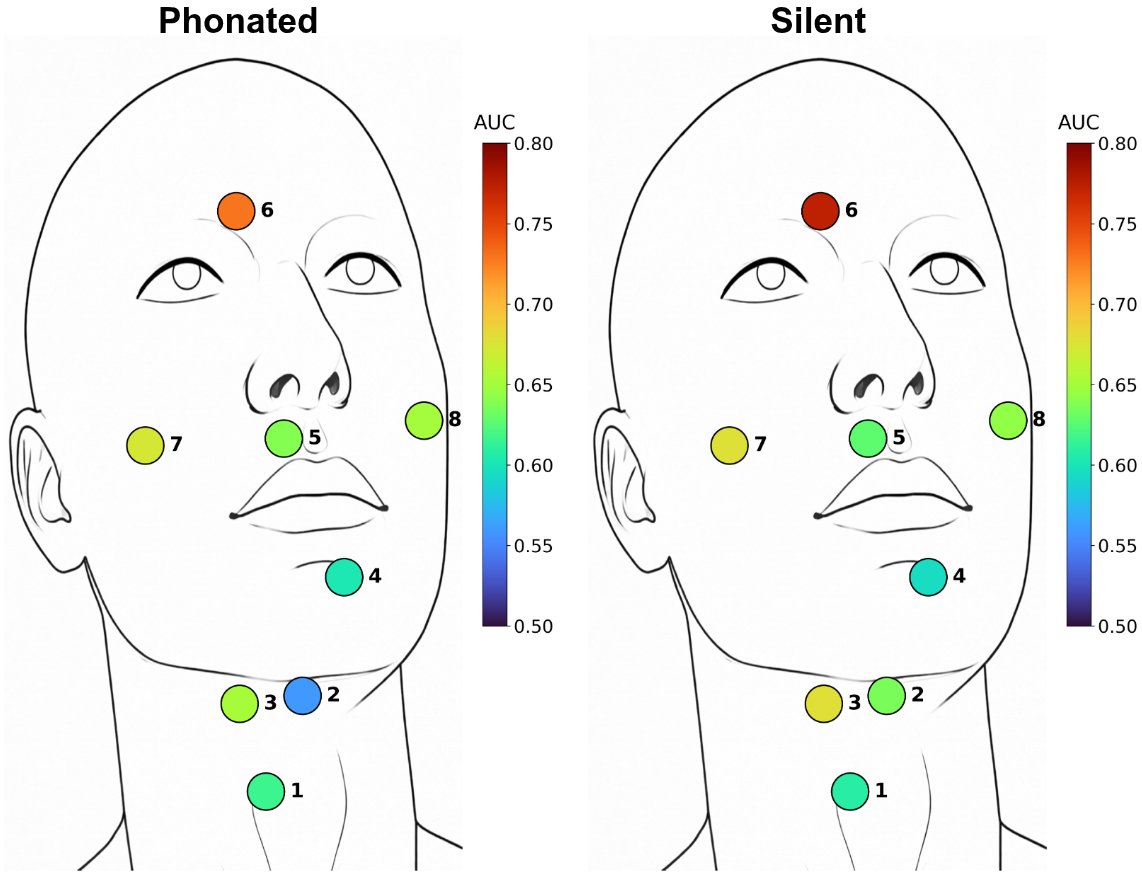}
  \caption{\textbf{Channel-wise decoding performance across articulation conditions (RQ2).} Topographic visualization of electrode-specific AUC for the phonated (left) and silent (right) conditions in Tasks 1 and 3. Each numbered marker corresponds to an EMG channel, and color indicates intra-subject AUC, with warmer colors reflecting higher discriminability.}
  \label{fig:ablation2}
\end{figure}

\subsection{RQ2: Affect decoding of phonated and silent speech}

Since affect information was found concentrated on the facial region, we hypothesized that changes in articulation manner would not cause severe performance regressions. Table~\ref{tab:silent_audible} shows the experimental results on Tasks 1 and 3, where we trained and validated our models in either phonated or silent trials. In particular, intra-mode performance appeared consistent with our earlier results and did not differ significantly between phonated and silent settings. We even observed a marginal improvement for silent speech of about 2 percentage points, particularly for BioCodec embeddings. This difference was even larger for inter-subject models, however, did not reach significance.

With respect to cross-mode performance, the intra-subject results indicate that training on phonated speech could robustly transfer to silent-speech settings, whereas the reverse was true only for the BioCodec embeddings, i.e., both structural and TD features showed regressions of 3 to 7 percentage points. This result holds promise for silent speech interfaces that could be trained on typical phonated speech without explicitly requiring silent-speech recordings. For the inter-subject setting, this effect was attenuated with no significant differences.

Figure~\ref{fig:boxplots} further compares intra-subject AUC between Task 1 and Task 3 across phonated and silent speaking conditions. When evaluated on all sentences (left panel), performance in the phonated condition remains largely unchanged between the two tasks, whereas the silent condition shows a clear increase in Task 3. This improvement is marginally significant (paired t-test, $p = 0.046$) and likely reflects increased familiarity with the task, particularly for silent articulation, which constitutes the less natural and more demanding mode for typical speakers. This familiarity effect also extends to the more challenging subset of repeated sentences (right panel), where silent performance again improves in Task 3 (albeit underpowered, $p = 0.252$), while phonated performance remains relatively unchanged. Overall, these findings suggest that task repetition primarily benefits silent EMG-based decoding, and that silent articulation performance is more sensitive to learning and adaptation effects than phonated speech.

We additionally performed the same interpretability analysis, and present channel-wise results in Figure~\ref{fig:ablation2}, for phonated and silent articulation conditions. Overall, E6 exhibited the highest discriminability in both settings, with increased performance in the silent condition, i.e., from 0.725 to 0.771 mean AUC. In contrast, submental and lower-neck channels display more variability, with some sites demonstrating improved AUC during silent speech production. This pattern suggests that while upper facial regions consistently encode affect-related effort, silent articulation may reflect increased reliance on articulatory muscle engagement in the absence of overt phonation. Such an effect is plausible given that silent articulation is a less familiar and less automatized motor behavior.


\begin{table}[t]
\centering
\caption{\textbf{Inter-subject} performance on frustration detection (Task 2) in terms of mean $\pm$ standard deviation of trial-wise Balanced Accuracy (BAC) and Area Under the Curve (AUC).}
\begin{tabular}{l l cc}
\toprule
\textbf{Modality} & \textbf{Features} & \textbf{BAC} & \textbf{AUC} \\
\midrule

\multirow{3}{*}{EMG}
& Structural  
& \underline{0.616} $\pm$ 0.003 
& 0.623 $\pm$ 0.005 \\
& TD-0   
& 0.527 $\pm$ 0.006 
& 0.518 $\pm$ 0.004 \\
& BioCodec  
& 0.595 $\pm$ 0.014 
& 0.630 $\pm$ 0.009 \\

\midrule

\multirow{2}{*}{Speech}
& eGeMAPS      
& 0.607 $\pm$ 0.002 
& \underline{0.679} $\pm$ 0.003 \\
& Vox-Profile  
& \textbf{0.670} $\pm$ 0.006 
& \textbf{0.743} $\pm$ 0.004 \\

\bottomrule
\end{tabular}
\label{tab:spontaneous}
\end{table}

\subsection{RQ3: EMG markers in spontaneous speech}

Having established the feasibility of EMG-based affect decoding in prompted trials, we proceeded to evaluate whether these models could generalize to spontaneous speech across speakers. In Table~\ref{tab:spontaneous}, we present the results regarding Task 2 of our experiment. Here, models were trained on controlled phonated and silent tasks (Tasks 1 and 3) from all but the held-out speaker, and test on the spontaneous (phonated) frustrated and polite trials of that speaker. We note that evaluation metrics are reported here on the individual-trial level to account for the variable number of spontaneous utterances across the 12 participants.

As shown in Table~\ref{tab:spontaneous}, speech-based models achieve the highest inter-subject performance on spontaneous frustration detection, with Vox-Profile embeddings yielding the best overall results (BAC = 0.670; AUC = 0.743), followed by the prosodic eGeMAPS feature set (AUC = 0.679). In contrast, EMG-based models demonstrate more moderate but consistently above-chance performance. BioCodec achieves the highest discriminability (AUC = 0.630), while structural features yield a comparable BAC of 0.616. TD performs lower overall in this spontaneous cross-speaker setting. These findings suggest that spontaneous speech provides richer acoustic cues for affect decoding, benefiting pretrained speech representations. At the same time, EMG models achieve performance comparable to handcrafted prosodic features (eGeMAPS) and exhibit improved balanced accuracy relative to the inter-subject controlled condition, indicating partial generalization from prompted to spontaneous affect expression.

\begin{figure}[t]
  \centering
  \includegraphics[width=\linewidth]{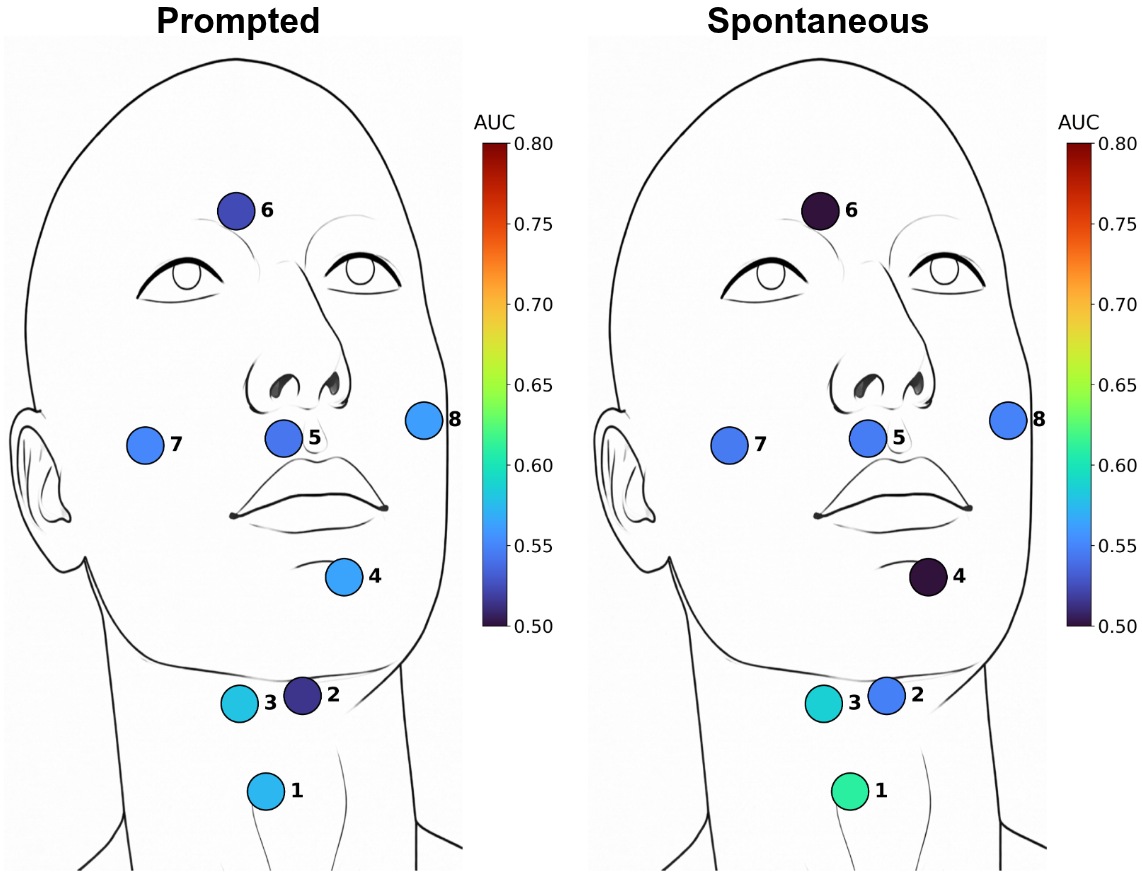}
  \caption{\textbf{Channel-wise decoding comparison between controlled and spontaneous settings (RQ3).} Topographic visualization of electrode-specific AUC for inter-subject performance in Tasks 1 and 3 (prompted, left) and Task 2 (spontaneous, right). Each numbered marker corresponds to an EMG channel, with warmer colors reflecting higher discriminability.}
  \label{fig:ablation3}
\end{figure}

Figure~\ref{fig:ablation3} further contrasts the spatial distribution of channel-wise inter-subject decoding performance between controlled (prompted) and spontaneous speech settings. In the prompted-phonated condition, performance is relatively homogeneous across facial sites, with moderate discriminability observed in perioral and cheek channels. In the spontaneous setting, overall AUC decreases for several frontal and mid-face channels while certain submental and lower facial sites exhibit comparatively enhanced performance. This redistribution suggests that affect-related neuromuscular patterns differ between controlled articulation and naturalistic speech, with spontaneous production potentially engaging more prosodic/acoustic elements, which could also explain the robust performance of the audio models in this task (Table~\ref{tab:spontaneous}). Overall, the results highlight a shift in spatial encoding of affect when moving from structured to ecologically valid speaking contexts.


\section{Conclusions} 

The presented findings support the view that affective modulation is embedded in broad facial motor execution rather than solely at the acoustic level. Extending prior EMG studies conducted in passive emotion paradigms~\cite{tan2016recognition,sato2013relationships}, we demonstrate that affective signatures persist within structured speech production and even during silent articulation. Interestingly, our ablation study indicated that cross-speaker transfer shifted discriminability from facial toward lower-neck channels, pointing to greater inter-speaker consistency in laryngeal and prosodic-related motor components. This aligns with findings from silent speech research showing substantial heterogeneity in facial EMG~\cite{golland2018affect}. Overall, while our models showed a moderate degree of generalizability, this should be assessed under the assumption of inherent heterogeneity in both muscle physiology~\cite{vigotsky2018interpreting} and speech emotion expression~\cite{van2023modelling}. 

Despite this variability, relatively simple EMG features were sufficient to capture affective patterns in most of our experimental settings, which could enable rich future work into model interpretability, specifically regarding the coupling between motor patterns to specific prosodic variations~\cite{busso2007interrelation}. Still, the competitive performance of BioCodec features in inter-subject, repeated-sentence, and cross-mode conditions suggests that learned embeddings may capture more transferable structure than handcrafted features in challenging inference scenarios. This comes despite the fact that this model was pretrained on non-speech EMG~\cite{avramidis2025neural} and applied in a zero-shot manner.

\textbf{Study limitations:\;} Several limitations should be acknowledged in the context of this study. The participant cohort was modest in size and demographically imbalanced, which would limit any population-level conclusions from our inter-subject performance. Furthermore, the decoded affective states (politeness and frustration) were experimentally prompted and induced, rather than occurring in ecological settings. The same applies to silent articulation, which was explicitly acted and inherited a learning curve for the participants (Figure~\ref{fig:boxplots}). Finally, it is important to note that our study by design could not disentangle articulatory modulation from co-activated facial expressions accompanying speech. Overall, these findings suggest that affect is not only audible but embodied in the neuromuscular processes underlying speech production, motivating future work in larger and more ecologically valid settings. 

\section{Acknowledgments}
The authors would like to thank the Bavarian Californian Technology Center (BaCaTec) for their financial support and for facilitating this collaboration. This work was also supported by the German Research Foundation (DFG) under the project "Silent Paralinguistics" (Grant No. 40301193).

\section{Generative AI Use Disclosure}

Generative AI tools were used in this study to assist with language polishing, manuscript editing, and assisting code implementations for analyses and visualizations relevant to this paper. These tools were not used for results generation, data analysis, data interpretation, or at any stage of data collection. All authors are fully aware of the extent of generative AI use in this work, take full responsibility for the content of the manuscript. 

\bibliographystyle{IEEEtran}
\bibliography{mybib}

\end{document}